\documentclass[11pts,onecolumn]{article}
\usepackage[dvips]{graphics}
\usepackage{epsfig}
\usepackage{hyperref}
\usepackage[centertags]{amsmath}
\usepackage{amsfonts}
\usepackage{amssymb}
\usepackage{amsthm}
\usepackage{newlfont}

\begin{document}
\title{An Economic analogy to Electrodynamics}

\author{Sanjay Dasari,\\
Malad Office, Mumbai Branch, JPMorgan India Pvt. Services Ltd;\\
Formerly, 
Electronics and Instrumentation and M.Sc(Hons)Economics,\\
BITS-Pilani Goa Campus, Goa-403726.\\
Anindya Kumar Biswas, Department of Physics;\\
North-Eastern Hill University, Mawkynroh-Umshing, Shillong-793022.\\
email:anindya@nehu.ac.in}
\date{\today}
\maketitle
\begin{abstract}
\noindent
In this article, we would like to find the laws of
electrodynamics in simple economic systems. We
identify the chief economic variables and parameters, scalar and
vector. We find laws of economics interms of these variables and 
parameters. The laws are similar in form to the laws of symmetric 
electrodynamics. Moreover,
we obtain Phillp's curve, recession, Black-Scholes formula, 
Supply-Demand line and Cobb-Douglas production function as
sample applications.
\end{abstract}


\begin{section}{Introduction}
''We want the most but cannot have it'' as a saying 
goes. The saying says about the problem of scarcity, 
talks about the problem of duality. The explicit and 
implicit statements of the saying is at the core of the 
branch of economics. Economics being a system, being a 
problem have mesmerised many, including physicists. Physicists 
have tried to comprehend the complexity of economics from time 
immemorial, starting from Copernicus, through Isaac Newton to 
Eugene Stanley\cite{{Jean1},{Jean2},{saslow}}.

\noindent
The question keeps coming, can we understand economics as simply
as mechanics\cite{pikle}? Can we comprehend force laws behind
economic developments as simply as four force laws in physics?
Though there are few interesting attempts\cite{{eco1},{eco2},{eco4}}, 
direct attacks to answer the questions probably are missing.

\noindent
How a system changes position, in a given environment, as 
a function of time is the study of mechanics. How a given 
charged environment dictates a charged particle has been 
the branch of electrodynamics dealing with. With electrodynamics 
came electric and magnetic fields, electric and magnetic charges. 
Though the magnetic charge is yet to be discovered, the dichotomy, or,
duality of electric and magnetic is at the heart of enormous amount 
of theoretical constructions. One such construction is symmetric 
electrodynamics due to Cabibo and Ferrari, \cite{Singleton}.

\noindent
In this article, we will refer to the easily available books on
electrodynamics, \cite{david} and economics,\cite{eco}, while trying
to separate, step by step, one kind of force law in action in
economics. We do this in the following way. First we describe the
Maxwell's equations of electrodynamics as well as continuity
equation and Lorentz force law in the section II. Then we introduce the chief
economic variables and formulate the correspondence of the
economic variables to the standard electrodynamic variables and
parameters in the section III and section IV respectively.  
Virtues of competition were estimated highly by pioneering Adam Smith\cite{Adam}. 
Competition flow, here, is one of the chief 
economic variables.
In the next step, in the section V, we verify how equations of electrodynamics
are holding good in economic systems. We also consider analogue of
materials in economics in the section VI. Potential formulation of electrodynamics
is a powerful solution technique. We will see how that too
descends down to us in economics in the section VII.

\noindent
Formulation of economics based on the analogy to Maxwellian 
electrodynamics is sufficient to account for money and scarcity. 
To account for capital and labour we require symmetric completion 
of Maxwellian electrodynamics. Capital and labour gets their due 
places as monopole and antimonopole got in symmetric electrodynamics, 
\cite{Singleton}.

\noindent
Unemployment, inflation of prices are day to day headache.
Recession was the first word of the song for the day to start with
until few years back. What is less heard that there is an empirical
graphical relation between inflation rate and unemployment rate,
in the short run. The name of the line is Phillip's curve, due originally in an 
alternative form to A. W. Phillips, \cite{Phillips}.  We
derive sort of Phillip's curve using the rules, describe the
recession also in the subsections VIII.1 and VIII.2 
respectively. Moreover, option trading (one type of booking
share) is something that makes the share market efficient. Pricing
of the option has been a long standing academic issue. F. Black
and M. Scholes were the first to, using intuition from Physics,
namely diffusion equation of heat, give a reasonable formula
\cite{bs1} for the call (and hence put) option. In this article, we
re-derive the Black-Scholes formula, visualising call option as
one component of profit flow rather than temperature, as a
particular case of more general class of feasible formulas in the 
subsection VIII.3.
Unobservable factor market volatility, too gets split up. As 
sample applications requiring capital-labour sector, we derive 
supply-demand line and Cobb-Douglas production function in the 
subsections VIII.4 and VIII.5 respectively. In an 
appendix, section XIII we describe how we can embed utility in this formulation.
We consider subtle points, broader outlook and conclusion in the sections 
IX, X, XI respectively.

\noindent
We will take India and Indian currency, Rupee, as a background for
our purpose of the paper. But the full content will be holding
true, if India and Indian currency are replaced globally, in this 
article, by any country and the corresponding currency.
\end{section}
\begin{section}{Maxwell's equations}
We recall that the basic variables of electrodynamics are electric
field, $\overrightarrow{E}$ and magnetic field,
$\overrightarrow{B}$. These two fields can exist without, can
generate in a medium or, can be produced by electric (magnetic) charge
density, $\rho_{e}(\rho_{m})$ and electric current density,
$\overrightarrow{j_{e}}(\overrightarrow{j_{m}})$. The relations, 
whenever relevant, between
electromagnetic fields and charge(s) (current(s)) in 
a vacuum (material medium) are fixed by permittivity constant, 
$\epsilon_{0}(\epsilon_{0}\epsilon_{r})$, and
permeability constant, $\mu_{0}(\mu_{0}\mu_{r})$. These six variables have an
interesting interrelationship. Moreover, the charge density and
current constrain each other through a constitutive relation. Let
us describe along that line in the paragraph to follow.

\noindent
The four equations of symmetric electrodynamics  are
as [\cite{david}, \cite{Singleton}]
\begin{equation}\label{e1}
\epsilon_{0}\nabla\cdot\overrightarrow{E}= \rho_{e}
\end{equation}
\begin{equation}\label{e2}
\nabla \times\overrightarrow{E}=-\frac{\partial}{\partial t
}\overrightarrow{B}-\mu_{0}\overrightarrow{j}_{m}
\end{equation}
\begin{equation}\label{e3}
\nabla\cdot\overrightarrow{B}=\mu_{0}\rho_{m}
\end{equation}
\begin{equation}\label{e4}
\nabla \times\overrightarrow{B}=\mu_{0}\epsilon_{0}
\frac{\partial}{\partial t }\overrightarrow{E}+
\mu_{0}\overrightarrow{j}_{e}
\end{equation}
The constitutive relation(s), called continuity equation(s), is
\begin{equation}\label{con}
\nabla\cdot\overrightarrow{j}_{e,m}+\frac{\partial}{\partial t}\rho_{e,m} = 0
\end{equation}
The force acting on a charge (electric+magnetic) distribution 
is given by the Lorentz Force Law
\begin{equation}\label{lor}
\overrightarrow{F}=\rho_{e}(\overrightarrow{E}+\overrightarrow{v}\times\overrightarrow{B})
    +\rho_{m}(\overrightarrow{B}-\frac{1}{c^2}\overrightarrow{v}\times\overrightarrow{E})
\end{equation}
\subsection{Potential}
Ala Singleton\cite{Singleton}, we write $\overrightarrow{E}$ and $\overrightarrow{B}$
interms of two four potentials as 
\begin{eqnarray}
\overrightarrow{E}= -\nabla\phi_{e}-\frac{\partial\overrightarrow{A} }{\partial t}
                                     -\nabla \times\overrightarrow{C},\\
\overrightarrow{B}= -\nabla\phi_{m}-\frac{\partial\overrightarrow{C}}{\partial t}
                                     +\nabla \times\overrightarrow{A}                                     
\end{eqnarray}
In the Coloumb gauge, the equations, (\ref{e1}-\ref{e4}), reduce to
\begin{equation}
 \nabla^{2}\phi_{e,m}=-\rho_{e,m}
\end{equation}
and the corresponding Poisson's equations for vector potentials.
\subsection{Duality}
The equations (\ref{e1}-\ref{e4},\ref{con}) are covariant whereas eq.(\ref{lor})
is invariant under the duality transformation
\begin{eqnarray}
\overrightarrow{E}^{'} =\overrightarrow{E} cos\alpha +c\overrightarrow{B}sin\alpha\\
c\overrightarrow{B}^{'}=-\overrightarrow{E}sin\alpha +c\overrightarrow{B}cos\alpha\\
c\rho_{e}^{'}=c\rho_{e}cos\alpha + \rho_{m} sin\alpha\\
\rho_{m}^{'}=-c\rho_{e}sin\alpha + \rho_{m} cos\alpha\\
c\overrightarrow{j}_{e}^{'}=c\overrightarrow{j}_{e}cos\alpha 
                           +\overrightarrow{j}_{m}sin\alpha\\
\overrightarrow{j}_{m}^{'}=-c\overrightarrow{j}_{e}sin\alpha 
                           +\overrightarrow{j}_{m}cos\alpha\\
c\phi_{e}^{'}= c\phi_{e} cos\alpha + \phi_{m} sin\alpha\\
\phi_{m}^{'}=-c\phi_{e} sin\alpha + \phi_{m}cos\alpha
\end{eqnarray}
Moreover, $c^{2}\phi_{e}^{2} +\phi_{m}^{2}$ remains invariant under duality 
transformation.
\end{section}

\begin{section}{Analogous economic variables} 
We denote the main
economic variables as follows:
\begin{itemize}
\item
competition flow as $\overrightarrow{c}$
\item
profit flow as $\overrightarrow{P}$
\item
money flow as $\overrightarrow{M}$
\item
money density, money per unit volume, as $n$
\item
Ambition of a person as $\overrightarrow{Am}$
\item
Price index desirable by a consumer as $Pi_{c}$
\item
Price index desirable by a supplier as $Pi_{s}$
\item
Choice flow of a consumer (supplier) as $\overrightarrow{Ch}_{c,s}$
\item
Economic power flow as $\overrightarrow{E_{p}}$
\item
Economic activity as $E_{a}$
\item
inverse of basic strength-scale of currency, at least for macro
economy, as $s_{0}$
\item
basic technical knowhow+political power, at least for macro
economy, as $k_{0}$
\item
human infrastructure as h
\item
capital density as $\rho_{K}$
\item
capital flow as $\overrightarrow{K}$
\end{itemize}

\end{section}
\begin{section}{Correspondence}
\begin{itemize}
\item
$\overrightarrow{E}\longleftrightarrow\overrightarrow{c}$
\item
$\overrightarrow{B}\longleftrightarrow\overrightarrow{P}$
\item
$\overrightarrow{j}_{e}\longleftrightarrow-\overrightarrow{M}$
\item
$\rho_{e}\longleftrightarrow -n$
\item
$\overrightarrow{j}_{m}\longleftrightarrow\overrightarrow{K}$
\item
$\rho_{m}\longleftrightarrow \rho_{K}$
\end{itemize}
\subsection{parameters}
\begin{itemize}
\item
$\epsilon_{0}\longleftrightarrow s_{0}$
\item
$\mu_{0}\longleftrightarrow k_{0}$
\item
$\epsilon_{0}\epsilon_{r}\longleftrightarrow s $
\item
$\mu_{0}\mu_{r}\longleftrightarrow k$
\item
$\sigma\longleftrightarrow$ h
\end{itemize}
\subsection{functions}
\begin{itemize}
\item
$\overrightarrow{v}\longleftrightarrow\overrightarrow{Am}$
\item
Scalar potential, $ \phi_{e,m}\longleftrightarrow -Pi_{c,s}$
\item
Vector potential,
$\overrightarrow{A}\longleftrightarrow-\overrightarrow{Ch}_{c}$
\item
Vector potential,
$\overrightarrow{C}\longleftrightarrow-\overrightarrow{Ch}_{s}$
\item
Poynting vector, $\overrightarrow{S}
=\frac{1}{\mu_{0}}\overrightarrow{E}\times\overrightarrow{B}\longleftrightarrow
\overrightarrow{E_{p}}$
\item
energy density $\longleftrightarrow E_{a}$
\item
$\sigma_{cross}$multiplied by power,P
 $\longleftrightarrow$ $<employment>$,\\ employment generation rate
\end{itemize}

\end{section}
\begin{section}{Analogy brought inside out}
\subsection{Maxwell's equations}
\begin{itemize}
\item
Excess liquidity stimulates economic activity i.e. generates
competition. Faraway from mints, activity drops to zero,
competition fizzles out.

\noindent
To understand it better, let us consider the following simple
situation, one has left a one rupee note on the road separating
two parts of a market, it will lead to a competition among the
onlookers to pick it up. Imagine, instead one lakh rupee note kept
on the road. It will lead to fiercer competition among the
onlookers. Not only that competition which is under way along the
road or, along either part of the market, will get a component
across the road. Hence money density in a place generates
divergence in competition flow and proportional. This is
proportional at least to the first approximation. Moreover,
competition points towards the money.

\noindent
Let us think the exactly same situation happening twenty five
years back. Then, one rupee note would have given the same
divergence in the competition flow as ten thousand rupees give
today. Within past twenty five years, rupee has gotten devalued by
huge amount. Hence, the proportionality factor $s_{0}$ stands for
the inverse of strength-scale of the currency.

\noindent
This sequence of arguments follow even if we consider not this
kind of free notes but constrained notes. We mean, the same kind
of situation will arise with the salary of an advertised job also.
We will be concerned in this paper with competition associated
with the constrained notes.

\noindent
Hence we deduce the first law analogous to the eq. (\ref{e1})
\begin{equation}\label{E1}
s_{0}\nabla.\overrightarrow{c}=-n
\end{equation}
\noindent
In this sense, money density is analogue of negative charge
density. Scarcity is analogue of positive charge. Scarcity density
is more like hole density than free positive charge density. Note
and scarcity, in equal magnitude form dipole. An arbitrary
distribution of note (scarcity) over space can be cast into the
form of multipole  expansion.

\noindent
In an organisation, when money is not flowing or, notes are
stationary there is no competition. This is like
$\overrightarrow{E}=0$ in a conductor.

\item
In general profit is a composite object composed of money, labor
etc. In the simplest cases profit is quantified as money gain. In
any exchange, positive profit of one is equal to, in magnitude,
the negative profit of the other. Hence, in any exchange, net
change in profit is zero. If there is no exchange, there is no
change in profit, either way. Hence, we have
\begin{equation}\label{E2}
\nabla.\overrightarrow{P}=0. 
\end{equation}
Let's consider an isolated primitive economy. Let's recall in 
this context the popular phrase ''hand to mouth''. The profit 
is starting from hand and ending at mouth. If we focuss locally 
on hand or, mouth, the above equation, (\ref{E2}), is not holding 
true. Hand is the source for profit, mouth is the sink. So we 
should introduce a source term on the right hand side of the 
equation, (\ref{E2}). Hand stands for labour. Extended hand, in 
economic parlance, is capital. As we graduate through closed 
to open economy, the appropriate definition, it appears to us, 
for labour is the non-signalling head+torso of an animate object.
Hands, feet, signalling mouth, traditional capital like tools, cabs,
houses, lands, old technologies and new technologies like software, 
internet, nano-tech or, stem-cell therapy . etc. comprise capital. 
With this definition of capital and labour, these two stand for 
positive and negative single pole sources for profit. The part 
of the profit lines starting from hand and ending at nearby 
head+torso are wage-lines, the rest other profit-lines are 
ending at somewhere else and termed as profit in traditional 
language. Hence, the eq.\ref{E2} is a special case of a general 
equation 
\begin{equation}\label{E22}
\nabla.\overrightarrow{P}=k_{0}\rho_{K}
\end{equation}
where, $\rho_{K}$ stands for capital density and $k_{0}$ is the 
basic sophistication scale of the capital. Capital is positive, labour 
is negative. Capital, labour can form dipole like machine and mechanic.
\item
Profit flow coming from retail chain sector leads local
businessmen to get united and protest. Protest is a form of
competition flow. We may note that this is what experienced in
pure diamagnetic phenomenon or, when a bar magnet is pushed
orthogonally towards a wire loop. Initial reactions to software
coming to India were also similar. This motivates us to write
\begin{equation}\label{E3}
\nabla \times\overrightarrow{c}=-\frac{\partial}{\partial t. 
}\overrightarrow{P}.
\end{equation}
This also indicates that Faraday's law boils down to Ricardo's
principle in economics.

\noindent
The eq.\ref{E3} takes the form, in the presence of capital,
\begin{equation}\label{E32}
\nabla \times\overrightarrow{c}=-\frac{\partial}{\partial t
}\overrightarrow{P}-k_{0}\overrightarrow{K}.
\end{equation}
\noindent
Just recall competition of biofarms about new patents, family members 
about a house, companies about contracts (say, gas, band-width, oil etc.).
Competition repels house to come to anyone in the family.
\item
Like magnetic field profit is also non-conservative field. If
there is no money, there is no profit. Circulation of notes gives
rise to profit. As money starts incoming more and more to a place,
profit also increases, say in a place, to some people more and
more. As money comes more, differences in money contents from
person to person, say, increase more. Rich becomes richer, poor
becomes poorer. This is a local consideration. 

\noindent
Let us think of the opposite limit, where there is no money flow
into a place. But if competition flow, say promotional competition
in a company, changes with time, like in some months of the year,
this leads to more spending, hence more profit circulation in the
local economy or, micro-economy. Product differentiation too leads
to circulation of profit in a local economy. These considerations
lead us to the relation
\begin{equation}\label{E4}
\nabla \times\overrightarrow{P}=s_{0}k_{0}
\frac{\partial}{\partial t }\overrightarrow{c}-
k_{0}\overrightarrow{M}
\end{equation}

\end{itemize}

\subsection{Continuity equation}
We know that no one creates (destroys) money, unless one is crazy.
The amount of money that enters (goes out) from one's pocket, or,
from one ATM, or, from one bank, in unit time is just equal to the
rate of change of money in that pocket or, ATM or, the bank. This
is just the continuity eq. (\ref{con}). \\
But there is an exception. Notes are destroyed or, generated at
the mint(s), leading to appreciation or, depreciation w.r.t. a
standard currency. \\
So the relation(\ref{con}) takes the following form, in case of economics
\begin{equation}
\nabla\cdot\overrightarrow{M}+\frac{\partial}{\partial t}n =
\frac{\partial}{\partial t}n_{p}
\end{equation}
where, $n_{p}$ is the amount of money being printed or, destroyed
in a mint.

\noindent
Applying the same kind of logic, we get to a continuity equation for
capital flow,
\begin{equation}
\nabla\cdot\overrightarrow{K}+\frac{\partial}{\partial t}\rho_{K} =
\frac{\partial}{\partial t}\rho_{K_{p}}
\end{equation}
\noindent
where, $\rho_{K_{p}}$ is the density of capital being created or, destroyed.

\subsection{Lorentz Force Law}
Let us imagine, competition has started flowing in a place, buy a
house or, buy sports goods or, buy a ticket for a show. A person
will respond or, not and if responds to what extent, depends on
how much money is there in his pocket. Whether a locality around
an ATM will respond or, not or, to what extent will depend on how
much notes are there at the ATM. Response varies directly also
with the appeal or, magnitude of the competition flow.  So the
force along the competition flow on a person or, a local society
around an ATM is proportional to the competition flow, to the
first approximation and the proportionality factor is money
density. The same thing occurs for a nation about a Federal bank,
in response to an oncoming competition flow. Here, we are meaning
by competition flow, social competition flow.

\noindent 
Let us consider an opposite situation. Reality sector
boom is coming onto a place, along the "third dimension". A person
will respond provided he has business ambition. The response will
be proportional to the money he owes. Once he responds this will
give sidewise pushes to the people around him, who might be
harbouring academic ambition only, on-setting competition along
the direction perpendicular to the person's ambition direction and
the profit flow direction.

\noindent 
Hence we heuristically come down to an equation of
economic force, which is exactly the same form as Lorentz force
law
\begin{equation}
\overrightarrow{F}=-n(\overrightarrow{c}+\overrightarrow{Am}\times\overrightarrow{P})
\end{equation}
Here, we observe that only competition flows cannot give a man
having scarcity, equilibrium but profit flows can. This is like
Earnshaw's theorem. Second part of the statement is like magnetic
confinement of charge. The same kind of logic in presence of capital 
leads us to 
\begin{equation}
\overrightarrow{F}=-n(\overrightarrow{c}+\overrightarrow{Am}\times\overrightarrow{P})
+\rho_{K} (\overrightarrow{P}-\frac{1}{c^{2}}\overrightarrow{Am}\times\overrightarrow{c})
\end{equation}
\noindent
Interestingly, the two terms with opposite sign, is the potential source of 
oscillation for any economy. In macroeconomy, we are familiar with observed 
business cycles.

\noindent 
Here, we also notice that two twins having the same
money, same ambition and subjected to the same competition and
profit flows, will feel the same force. But depending on their
accumulated entrepreneul skills their venture accelerations will
be different. For example, one will set-up a cyber cafe much
earlier than other, if the first one has software and little bit
management training whereas the second one does not have that
skill set. Hence economic inertial mass of a person is reciprocal
of the number of his entrepreneul skills. We denote from hereon,
\begin{itemize}
\item
economic inertial mass= $M_{e}$
\item
Number of skills= $N_{es}$
\end{itemize}
The same story will follow for two twin companies or, two twin
countries. Hence we have the following identification
\begin{itemize}
\item
$M_{e}=\frac{1}{N_{es}}$
\end{itemize}
\noindent
In the next section, we will discuss analogue 
of material and conductivity, 
restricting ourselves to the electric charge sector.
\end{section}
\begin{section}{Material}
Let us think that competition flow is oncoming to a
place. This will create money accumulation among some and scarcity
among others, giving rise to something like polarisation, bound
money density at the surface of the society and at the volume. As
a consequence, net competition flow will be different from the
external competition flow. For weakly responsive society,
polarisation vector will be equal to
$s_{0}R_{c}\overrightarrow{c}$. $R_{c}$ is the measure of the
response of the society. $\overrightarrow{c}$ refers to the net
competition flow in the society. The equation (\ref{E1}) will get
modified to
\begin{equation}
\nabla.s\overrightarrow{c}=-n.
\end{equation}
$n$ refers to external money density.
$s=s_{r}s_{0}=s_{0}(1+R_{c})$.

\noindent 
Similarly, profit flow leads to bound surface and volume
circulation of notes. This results in the net profit flow
differing from the external profit flow vector. This leads to a
relation modified from the equation (\ref{E2})
\begin{equation}
\nabla.k\overrightarrow{P}=0
\end{equation}
where, $k=k_{0}k_{r}=k_{0}(1+R_{p})$.\\
\noindent 
Probably, $s$, $k$ span a two dimensional plane.
Presumably, existence of black market is an example of $s$, $k$
being both
negative\cite{meta}.\\
Profit and competition flows both polarize.
\subsection{Conductivity}
Sometimes economy is conducive. Competition vector is proportional
to money flow vector or, liquidity just like in conductor,
\begin{equation}
\overrightarrow{j}= \sigma \overrightarrow{E}
\end{equation}
Proportionality factor, h, in economic system, like conductivity,
is a measure of the quality of the human infrastructure of the
company. So we have here the following rule
\begin{equation}\label{human}
\overrightarrow{M}=-h\overrightarrow{c}
\end{equation}
In highly efficient ($h\rightarrow \infty$) organisation, internal
competition is zero always, which is like in metal ($\sigma
\rightarrow \infty$). $h$ can stand for $Human Capital$, \cite{Human}.
\end{section}
\begin{section}{Potential Formulation} 
To show the form of the scalar
potential, let us notice the following,
\begin{equation}
\overrightarrow{c}=-\nabla (-Pi_{c}), \overrightarrow{P}=-\nabla (-Pi_{s});
\end{equation}
implies
\begin{equation}
\nabla^{2} Pi_{c}= -n\frac{1}{s_{0}}, \nabla^{2} Pi_{s}= -k_{0}\rho_{K}.
\end{equation}
As money density increases, Price-index also increases, we see
inflation.

\noindent 
Price index over space and time is determined by two
considerations
\begin{itemize}
\item
Prices and consumption ratios of various items at a place at a
given time.
\item
Prices and consumption ratios of items at another time and/or at
another place, compared to the base prices and consumption ratios.
\end{itemize}
The prices and consumption ratios of items change continuously
over the space and time.

\noindent 
Hence, Price index, $Pi_{c,s}$, change continuously over space
and time. So, Price index, $Pi_{c,s}$, is analogous to scalar potential,
$\phi_{e,m}$. The first consideration sets a fixed reference value to the
price-index for all other places at that time as well as for all
other times.  A relevant fact worth mentioning in this context is
that gas index in U.S. is based on the price of gas at a point
where majority of the gas pipelines intersect.

\noindent
To show the form of the vector potential, let us notice
the following,
\begin{equation}
\nabla^{2}\overrightarrow{Ch}_{c}=-k_{0}\overrightarrow{M}, 
\nabla^{2}\overrightarrow{Ch}_{s}=k_{0}\overrightarrow{K}
\end{equation}
wherever, choice flow is divergence less. This continues to be as
long as there is no will.

\noindent
Hence, $\overrightarrow{Ch}_{c}$ is in the same direction as
$\overrightarrow{M}$, 
as  $\overrightarrow{Ch}_{s}$ is in the opposite direction as
that of $\overrightarrow{K}$ which is our experience.

\noindent 
Moreover, $(Pi_{c,s},\overrightarrow{Ch}_{c,s})$ can be combined
into a four vector.  Ambition, $\overrightarrow{Am}$, multiplied
by Price index can be choice. Maximum Ambition is determined by
the velocity of light and in fact, may be taken as velocity of
light. We would like to move in any direction with the magnitude
of velocity of light, c, given chance. Therefore it's quite
plausible to write
\begin{equation}
\overrightarrow{Ch^{'}}=\frac{\overrightarrow{Ch}-\overrightarrow{Am}Pi}{\sqrt{1-\frac{Am^{2}}{c^{2}}}}
\end{equation}

\subsection{Duality}
The equations (\ref{E1},\ref{E22},\ref{E32},\ref{E4}) are covariant whereas eq.(27) is invariant under the 
duality transformation
\begin{eqnarray}
\overrightarrow{c}^{'}=\overrightarrow{c}cos\alpha +c\overrightarrow{P}sin\alpha, \\
c\overrightarrow{P}^{'}=-\overrightarrow{c}sin\alpha +c\overrightarrow{P}cos\alpha, \\
-cn^{'}=-cncos\alpha+\rho_{K} sin\alpha,\\
\rho^{'}_{K}=cn sin\alpha+\rho_{K} cos\alpha,\\
-c\overrightarrow{M}^{'}=-c\overrightarrow{M}cos\alpha +\overrightarrow{K}sin\alpha, \\
\overrightarrow{K}^{'}=c\overrightarrow{M}sin\alpha +\overrightarrow{K}cos\alpha, \\
cPi^{'}_{c}=cPi_{c} cos\alpha+Pi_{s} sin\alpha,\\
Pi^{'}_{s}=-cPi_{c} sin\alpha+Pi_{s} cos\alpha
\end{eqnarray}
Moreover, $c^{2}Pi_{c}^{2} +Pi_{s}^{2}$ remains invariant under duality 
transformation.
\end{section}

\begin{section}{Application}
In the next three subsections we restrict ourselves to the electric charge 
sector solely.
\subsection{Phillip's curve} 
We know, in economics, Inflation rate, $\Pi$, is defined as
\begin{equation}
\Pi=\frac{d}{dt}{lnPi}.
\end{equation}
Since,
\begin{equation}
\phi_{e}\leftrightarrow Pi_{c}, \nonumber
\end{equation}
\begin{equation}\label{infla1}
 \frac{d}{dt}{ln\phi_{e}}\leftrightarrow \Pi 
\end{equation}
or, time derivative of logarithm of scalar potential is expected
to show features of economic inflation. To proceed along that
line, we note from the theory of radiation in electrodynamics,
\begin{equation}\label{infla2}
\frac{d}{dt}{ln\phi_{e}}=\omega,
\end{equation}
for electric dipole radiation, whereas, the total power radiated
by the dipole is given by
\begin{equation}
<P>=constant \quad \omega^{4}
\end{equation}
Hence,
\begin{equation}
\frac{d}{dt}{ln\phi_{e}}\sim <P>^{\frac{1}{4}}
\end{equation}
\noindent Here we recall that when an electromagnetic radiation
falls on a medium, three processes occur. For low energy,
photoelectric effect is the dominant process. As the energy
increases of the infalling radiation, Compton scattering starts
becoming important. At still higher energy, pair production takes
over. For the photoelectric effect, cross-section,
$\sigma_{cross}$, or, probability for the process to occur
\begin{equation}
\sigma_{cross} \sim \frac{1}{\omega^{\frac{7}{2}}}
\end{equation}
\noindent
Photoelectric effect is producing free electrons at the cost of
work-function. This phenomenon is exactly similar to employment
generation from the pool of unemployed youth at the cost of lump
sum money. In India, this is like giving one-time small money/loan
to buy say an auto/a cab to an unemployed young man and making him
self-employed. 
Hence, employment generation rate, 
denoted as $<employment>$ is the analogue of total transition rate, 
$<P>\sigma_{cross}$. But 
\begin{equation}
<P>\sigma_{cross} \sim \omega^{\frac{1}{2}}.
\end{equation}
Or, 
\begin{equation}
 <employment> \leftrightarrow \omega^{\frac{1}{2}}.
\end{equation}
\noindent
Again we know, product of employment generation rate and
unemployment generation rate is constant, because the two
processes occur in mutually exclusive sectors, influencing each
other in extreme cases, viz. percolation of software jobs to
mechanical and clerical sectors. In other words, 
\begin{equation}
<employment> <unemployment>=constant.
\end{equation}
This implies 
\begin{eqnarray}
<unemployment> \leftrightarrow \frac{1}{\omega^{\frac{1}{2}}},\\
<unemployment>^{-2} \leftrightarrow \omega .
\end{eqnarray}
At the same time, equations, (\ref{infla1},\ref{infla2}), together mean 
for the Inflation rate, $\Pi$
\begin{equation}
 \Pi \leftrightarrow \omega.
\end{equation}
Since two economic quantities, $\Pi$ and $<unemployment>^{-2}$, are 
analogue of $\omega$, these two must be proportional to each other.
In other words, for 
the low scale economic activity inflow,
\begin{equation}
\Pi\sim\frac{1}{<unemployment>^2}.
\end{equation}
\noindent 
This is nothing but Phillip's curve, qualitatively.
\noindent
Moreover, we note that the ongoing analysis in this section is 
reminiscent of finding 
relation between variables in physics, using dimensional analysis.

\noindent 
On the other hand, Compton scattering is pumping money in risky
assets. Pair production is like bringing an woman to work place at
the cost of a vacancy at the household cores. As a result, 
in the domain where Compton
scattering becomes important\cite{bjorken},
\begin{equation}
\sigma_{cross} \sim\frac{1}{\omega}ln\omega.
\end{equation}
Then
\begin{equation}
\Pi \sim\frac{1}{<unemployment>^{\frac{1}{3}}}.
\end{equation}
apart from the slowly varying scale-dependent logarithmic part.

\noindent 
Hence, in the scale of economic activity inflow,
$|\overrightarrow{E_{p}}|$ where, Compton scattering-type of
phenomenon becomes important compared to photoelectric type, we
get sudden increase of inflation with unemployment. This is
stagflation. This is stagflation with scale-dependence setting in.
\noindent 
If one is interested in total absorption cross-section,
one can look in \cite{kaplan} as well as in \cite{heitler} and
surmise about the details of the ensuing Inflation vs unemployment
curve.

\noindent
For small range of time, we get one kind of society, say one 
kind of regime, throughout the world. This is like one kind of 
material, say lead, throughout the whole space. For this situation, 
if we consider inflation rate in the horizontal direction and 
unemployment rate in the vertical direction, for the same dipole, for 
one ''$\omega$'' range, say in $\Delta\omega$, for $\Delta$inflation rate 
we get one kind of $\Delta$unemployment rate. But let us take many regimes, 
(long range), or, correspondingly many materials, say copper, zinc, lead etc.
For those for a given $\Delta$inflation rate we will have a series of 
$\Delta$unemployment rates, many non-smooth, say having discrete 
transitions from photoelectric to Compton . As a result, weighted sum of 
$\Delta$unemployment rates in one $\Delta$inflation rate will not join 
, in general, with that of the neighbouring 
$\Delta$inflation rate. Consequently, we do not get a smooth curve in the 
long range. In other words, the empirical relation, Phillip's curve, exists 
only for short run.

\subsection{Recession}
A Recessing phase corresponds to one inertial frame for a
macro-economy. The recessing inertial frame has lower ambition,
$|\overrightarrow{A_{m}}|$, with respect to that of an almost
contemporary macro-economy. Going to the recessing frame occurs
due to saturations of collective biological activities of the
society attached with the macro-economy.

\noindent
The inertial frame's ambition corresponding to the
macro-economy, can be thought as group ambition of the society.

\noindent 
As a result we see in the recessing phase, lower price
index, lower choice flow, hence lower consumption. This gets
manifest through deflation, unemployment.

\noindent 
Since $\nabla.\overrightarrow{Ch}$ is not Lorentz
invariant, $\nabla.\overrightarrow{Ch}\neq 0$ in the recessing
phase. This is like at mint $\nabla.\overrightarrow{M}\neq 0$.
That implies number of choice lines striking a populace from one
side is less than the number of lines leaving the populace in the
other side. That means human will is setting in and populace is
not spending to the brim. That is change in consumption pattern of
commodities as well as that of prices at each place with time.
This in turn will lead to lesser and lesser production and more
and more unemployment.
\subsection{Black-Scholes formula}
Let us suppose that we have gone to the stock-market armed with
the set of equations we have heuristically gotten and embark on
analysing the share trading. Moreover, let us focus on profit
attached with call option. Then the instantaneous profit is call
option value for someone having a share and writing a call option
for that share. Now let us try to find the value. Let us guide
ourselves by the thread of physical considerations of Black and
Scholes as appears in the first few pages of the
reference\cite{bs1}.

\noindent
As long as $\overrightarrow{E}$ which is analogue of competition
flow, $\overrightarrow{n}$, is constant or, slowly changing with
time, Maxwell's last two equations with the Ohm's law yields
\begin{equation}\label{e:bs}
\nabla^{2}\overrightarrow{B}=\mu_{0}\sigma\frac{\partial\overrightarrow{B}}{\partial
t}
\end{equation}
In terms of dimensionless length variables, this equation
(\ref{e:bs}) appears as
\begin{equation}
\frac{\partial\overrightarrow{B}}{\partial t}=\mu_{0}\sigma v^{2}
\nabla^{2}\overrightarrow{B},
\end{equation}
where, $|v|$ is the drift speed in the medium. Translating to
economic system by our dictionary and restricting us to the
variation of $\overrightarrow{P}$ along the third dimension, $x$,
say in the stock market, we get
\begin{equation}\label{black}
\frac{\partial P_{i}(x,t)}{\partial
t}=k_{0}h|\overrightarrow{Am}|^{2}
\frac{\partial^{2}P_{i}(x,t)}{\partial x^{2}}.
\end{equation}
where, for $i=1,2,3$, $P_{i}$ means  $P_{x}, P_{y}, P_{z}$.
Writing, $\tau=T-t$ and further doing the identification
\begin{itemize}
\item
implied volatility, $\sigma =\sqrt{2k_{0}h}|\overrightarrow{Am}|$
\item
$P_{i}=C(S,t)e^{r\tau}=u$ is the profit at time T, corresponding
to option trading at time t. $C(S,t)$ is the value of the option
when it is traded at time t. $C(S,T)=max(S-K,0)$
\end{itemize}
we get from the equation (\ref{black}) Black-Scholes differential
equation as given in the reference\cite{bs2},
\begin{equation}\label{black1}
\frac{\partial u(x,\tau)}{\partial \tau}=\frac{\sigma^{2}}{2}
\frac{\partial^{2}u(x,\tau)}{\partial x^{2}}.
\end{equation}
At this point let us do some more dimensional considerations: in
Option trading, relevant independent variables are
\begin{itemize}
\item
Current stock price at time $t = S$
\item
Strike price or, agreed upon price of the stock at the expiry i.e.
at time T is K
\item
Risk less interest rate is r (per year)
\item
Implied volatility in the stock price at time T is $\sigma$ where,
$\sigma^2$ has the dimension of time inverse (per year).
\end{itemize}
One way to combine these variables to get a dimensionless variable
$x$ is to write $x=ln\frac{S}{K} +(r-\frac{\sigma^2}{2})\tau$.
Once this is done, the straightforward solution of the equation
(\ref{black1}) yields the price of the call
option\cite{{bs1},{bs2}},
\begin{equation}
C(S,t)= SN(d_{1})-Ke^{-r(T-t)}N(d_{2})
\end{equation}
where,
\begin{eqnarray}
d_{1}=\frac{ln(\frac{S}{K})+(r+\frac{\sigma^2}{2})(T-t)}{\sigma\sqrt{T-t}},\nonumber\\
d_{2}=d_{1}-\sigma\sqrt{T-t},\nonumber\\
N(d)=\frac{1}{2\pi\sigma^2}\int_{\infty}^{d}
 dxe^{-\frac{x^2}{2\sigma^2}}\nonumber.
\end{eqnarray}
\subsection{Supply-Demand line}
Roughly, price index, $Pi_{c}$, desirable by a consumer, is given for 
two commodities system as 
\begin{equation}
Pi_{c} = 100(\frac{x}{X}.\frac{y}{y+q} + \frac{p}{P}.\frac{q}{y+q})=
 100\frac{\frac{x}{X}y+\frac{p}{P}q}{y+q}
\end{equation}
where, prices of two commodities are $x$ and $p$; whereas the quantities 
are $y$ and $q$ respectively. Prices of two commodities are $X$ and $P$ in the 
base year.

\noindent
Let us now take resort to Citrus Paribus condition i.e. assume that the 
price and the consumption of the second commodity remain fixed. Moreover, 
let's also assume that the note density and the strength scale of 
currency, $s$, remaining the same all over the space keeping price-index 
unchanging, courtsey eq.(28). Denoting the fixed price-index by $a$, 
we derive from eq.(64),
\begin{equation}
 x= X\frac{a}{100}+X\frac{q(\frac{a}{100}-\frac{p}{P})}{y}.
\end{equation}
So the plot between the price, $x$ and the quantity of the 
commodity number one, $y$,  is 
a hyperbola. We know, the middle part of the hyperbola is a straight line 
approximately. Hence, we have derived the equilibrium demand curve. By varying 
$a$, we get the series of demand lines.

\noindent
The object under bargain is duality invariant. The addition of squares of 
price indices as desirable by consumer and supplier is constant. Hence, for
a price-value of an object, addition of the squares of the quantity desirable 
for consumer and supplier is constant. Hence, supply-line has opposite 
orientation with respect to the demand-line. These two lines intersect at 
a point, called self-dual point. The price function, price per unit quantity, 
or, the supply-demand line has the shape of simple harmonic oscillator potential 
function. The self-dual point is the minimum of the potential function. Moreover, 
at the self-dual point the market clears. The self-dual point is where both the 
price indices agree and any one of these price index at that point is the 
familiar price index. On the other hand, the shape of the price function immediately 
entails possibilty of an oscillation(s) in the price, in the long run.

\subsection{Cobb-Douglas production function}
Production is attaching life to a substrate called material. Capital and 
labour combine to give life to the substrate to make it product. This is 
sort of reaction in which 
\begin{itemize}
\item 
capital life-time lost + labour life-time lost$\rightarrow$ product life-time 
\end{itemize}
This looks like 
\begin{itemize}
\item 
$\alpha$(unit capital life-time lost) + $\beta$(unit labour life-time lost) 
$\rightarrow$ product life-time 
\end{itemize}
Moreover, this appears like
\begin{itemize}
\item 
$\alpha$ A + $\beta$ B
$\rightarrow$ C
\end{itemize}
Now, the rate of production, $r$, is given by chemical kinetics as 
\begin{equation}
 r=A^{\alpha}B^{\beta},
\end{equation}
where, for a complicated multi-step process, stochiometric coefficients 
$\alpha$, $\beta$, can be any positive rational number. Chemical kinetics 
is nothing but electromagnetic forces in action. Using our dictionary and 
presence of duality symmetry implies that the rate of production, $q$, will 
have the form
\begin{equation}
 q=K^{\alpha} L^{\beta}.
\end{equation}
This is the Cobb-Douglas production function\cite{Cobb}.
\end{section}

\begin{section}{Points} 
Here we touch on some delicate issues. 
Competition flow
in this letter is separate from "pure arbitrage flow" just like
profit is more than money gain. We can think of three dimensional
vector spaces, locally composed of two dimensional plane and a
"third dimension". For a company, the "third dimension" is
hierarchy. In the stock market, the "third dimension" is the
"share" direction as we have explained in the previous subsection.
Normally, the "third dimension" is the third dimension,
communication is being made along that electrically or,
electromagnetically i.e. by land line or, satellite.

\noindent
Though appear distinct, profit flow and competition flow are close by 
in our formulation.
Profit making itself has a competitive spree within. 
Profit flow can be easily seen as a flow and can be measured. From 
the profit flow, one can, in principle, then calculate competition 
flow using relativistic transformations, \cite{david}, for many cases.
In cetain other situations, one can measure competition flow by 
measuring the difference 
of price-index as desirable by a consumer, $Pi_{c}$, at two places
and dividing by the distance between the two places.

\noindent
Moreover, any subject has three aspects, theoretical, descriptive and applied. 
Our analysis 
is theoretical unlike many other approaches to economics from physics 
which are applied in nature. Our analysis is local. It has the possibility 
of treating spatial inflation as well as spatio-temporal 
inflation, in particular and spatio-temporal economic phenomena, in general.
\end{section}
\begin{section}{Outlook}
Naively, one tends to wonder whether the topological
considerations in mathematical economics can be related to
magnetic topologies. Similarly many topics in economics,
elementary as well as advanced are expected to be described by
electrodynamics using the dictionary introduced in this paper. 
\noindent
One can take a straightforward route also. Consider the 
eqn.s(\ref{E1},\ref{E22},\ref{E32},\ref{E4})
as the rules of economics, measure the variables and the
parameters discussed, say, knowing $Human Capital$ ala economists,
one can try to measure competition flow using eq.\ref{human}, and
therefrom try to explain as many economic empirical relations as
possible. 
\end{section}
\begin{section}{Conclusion}
We have given an alternative formulation of economics. The rules
of the formulation are equations (\ref{E1},\ref{E22},\ref{E32},\ref{E4}). 
The variables are as mentioned, e.g. profit flow, competition flow, money flow,
constrained note density, capital etc. These rules are analogue of
symmetric Maxwell's equations. Moreover, we have obtained continuity
equation, force rule, inertial mass for an economic system and an
operational definition of $Human Capital$. We have constructed a
4-potential formulation. Using the model we get Phillip's curve,
describe stagflation, recession. Dwelling on stock-market we
recover Call option function. We have gotten a scenario where,
unobservable market volatility can be made observable if we can
measure the drift ambition of sort-sellers. 
We have made the known duality in economics more rigorous. Here, 
note and  capital are dual to each other, supply-demand line emerges 
as dual aspect of the same line, something as simple as 
Cobb-Douglas production function also gets 
a heuristic derivation via duality. Besides, 
we have pointed to few avenues, amidst many, 
along which this approach can be explored
further.
\end{section}
\begin{section}{acknowledgment}
To the best of our knowledge, the material covered
in this manuscript was not dealt with anywhere else. We would like 
to thank many people for discussions, comments and suggestions. 
\end{section}

\begin{section}{Appendix}
Here, to elucidate the concept of marginal utility, let us recall the 
prototype example of three oranges eaten one after another, thereby 
reducing the demand for a consecutive orange.

\noindent
Marginal utility(MU) is price. Hence, total utility(TU), is price index, 
magnitude-wise, once these are measured with money.
\begin{equation}
 MU=\frac{dTU}{dQ}
\end{equation}
But marginal utility and price are not identical. One is subjective and 
another one is objective. Hence, we go one step further and define marginal 
utility as $i$ times price, total utility as $i$ times price-index. Therefore, 
square of choice flow plus square of total utility is constant, as ambition 
changes.
\begin{equation}
 \overrightarrow{Ch}^{2} +TU^{2} = constant
\end{equation}
\noindent
Consequently, total utility is maximumwhen the choice flow has magnitude zero.
Keeping these in mind, let's analyse the example of oranges (but this time many 
oranges). Initially, choice is in the direction of eating orange and of large 
magnitude. As one by one, orange is being eaten, choice flow vector gets reduced
in length, as well as rotated. At one point, ''no more, please!'', choice 
becomes zero or, length of choice flow vector becomes zero.After thatif one 
is pushed to eat she/he starts feeling nauseatic, choice flow vector has 
reversed its direction and increasing its magnitude with each additional 
orange being swallowed. So with respect to the maximum value of total 
utility at a particular number of oranges ($Q_{m}$, say), (or, food set), 
the total utility reduces in both side as we increase or, decrease the number 
from $Q_{m}$. This is the standard total utility versus quantity graph found in 
economics book.
\end{section}

\end{document}